\documentstyle[12pt,epsfig]{article}
\newlength{\titlesep}
\newlength{\authorsep}
\makeatletter

\def\fnum@figure{FIG.~\thefigure}

\newcounter{figureparent}
\@addtoreset{figure}{figureparent}

\newcounter{eqnparent}
\@addtoreset{equation}{eqnparent}

\renewcommand{\abstract}{\if@twocolumn
  \section*{Abstract}
  \else
  \begin{center}
    {\bf Abstract\vspace{-.5em}\vspace{0pt}}
  \end{center}
  \quotation
  \fi}
\renewcommand{\endabstract}{\if@twocolumn\else\endquotation\fi}

\newcommand{\thismonth}{\ifcase\month\or
 January\or February\or March\or April\or May\or June\or
 July\or August\or September\or October\or November\or December\fi
 \space \number\year}

\makeatother

\newcommand{\newc}{\newcommand}
\newc\eg{{\it {e.g.}}}	\newc\vs{{\it {vs.}}}	\newc\etal{{\it {et al.}}}

\newc{\mhalf}{m_{1/2}}      \newc{\mzero}{m_0}

\newc\bsgamma{b\rightarrow s\gamma}
\newc\brbsgamma{BR(B\rightarrow X_s\gamma)}

\newc{\tanb}{\tan\beta}
\newc{\azero}{A_0}
\newc{\at}{A_t} \newc{\abot}{A_b} \newc{\atau}{A_\tau}
\newc{\bmu}{B\mu}           \newc{\sgn}{{\rm sgn}}
\newc{\mone}{M_1}           \newc{\mtwo}{M_2}

\newc{\bino}{\widetilde B}              \newc{\wino}{\widetilde W_3}
\newc{\higgsinob}{{\widetilde H}^0_b}   \newc{\higgsinot}{{\widetilde H}^0_t}

\newc{\mw}{m_{\rm W}}
\newc\msusy{M_{\rm SUSY}}
\newc{\mplanck}{M_{\rm P}}

\newc{\mub}{\overline{\mu}_{\rm b}}	\newc{\muw}{\overline{\mu}_{\rm W}}
\newc{\mususy}{\overline{\mu}_{\rm SUSY}}

\newc{\Ci}{C_i}	\newc{\Cip}{C_i^{\prime}}

\newc{\deltadll}{\delta^d_{LL}}	\newc{\deltadlr}{\delta^d_{LR}}
\newc{\deltadrl}{\delta^d_{RL}}	\newc{\deltadrr}{\delta^d_{RR}}

\newc{\abund}{\Omega h^2}
\newc{\abundchi}{\Omega_\chi h^2}
\newc{\rhocrit}{\rho_{crit}}
\newc{\rhochi}{\rho_{\chi}}

\newc{\xf}{x_f}
\newc{\jxf}{J({\xf})}
\newc{\VEV}[1]{\langle #1 \rangle}

\newc{\ra}{\rightarrow}
\newc{\beq}{\begin{equation}}
\newc{\eeq}{\end{equation}}
\newc{\bea}{\begin{eqnarray}}
\newc{\eea}{\end{eqnarray}}

\newcommand\lsim{\mathrel{\rlap{\lower4pt\hbox{\hskip1pt$\sim$}}
    \raise1pt\hbox{$<$}}}
\newcommand\gsim{\mathrel{\rlap{\lower4pt\hbox{\hskip1pt$\sim$}}
    \raise1pt\hbox{$>$}}}

\newcommand{\Rn}[1]{{\uppercase\expandafter{\romannumeral#1}}}

\newcommand{\epsfile}[1]{\relax}

\begin{document}
\pagestyle{empty}

\begin{center}
{\Large\bf
$B \to X_s \gamma$ in minimal supersymmetric standard model
 with general flavor mixing
\footnote{
Talk given by K.~Okumura at the 10th International
Conference on Supersymmetry and Unification of Fundamental
Interactions (SUSY02)  June 17-23, 2002, DESY Hamburg.
This talk is based on reference~\cite{or1}.}}\\
\vspace*{1em}
{\bf
Ken-ichi Okumura 
and Leszek Roszkowski
}\\
\vspace*{0.5em}
{\it 
Department of Physics, Lancaster University,
LANCASTER LA1 4YB, England
}\\
\end{center}
\begin{abstract}
\small
We present the results of our analysis on $\bsgamma$
 in the minimal supersymmetric standard model
 with a general flavor mixing in the squark sector.
We identify the next--to--leading order
 focusing effect, which considerably reduces
 the gluino contribution to $\bsgamma$ relative to the leading order calculation.
We illustrate that $\brbsgamma$ is still very sensitive
 to the flavor mixing parameters, $\deltadll$ and $\deltadlr$ so that
the lower bound for $\mhalf$ obtained in the minimal flavor violation 
can be easily removed even at $\mu<0$, 
while it is rather stable against $\deltadrr$ and $\deltadrl$.
\end{abstract}

\vspace{-1.em}
\small
\noindent
\section{ Introduction}

Radiative decays of B--mesons
 provide a unique opportunity for exploring physics beyond the SM.
In the SM, a leading contribution is from a virtual
 $W$--top (charm/up) loop accompanied by a real photon
 with GIM suppression.
New physics effects, which are also loop induced,
 could be same order if mass scale of the new physics is not far from
that of $W$ and top-quark. This is exactly the case of softly broken 
low--energy supersymmetry (SUSY).

At early stage of the investigation,
 $\bsgamma$ was expected to provide a clear signature of
 supersymmetry~\cite{bbmr90}, however, experimental and theoretical
 progress have  eventually revealed that there remains little room
 for a new physics contribution.
The world average 
of the branching ratio has reached an accuracy of $10$ \% level,
\beq
\brbsgamma = ( 3.41 \pm 0.36 )\times 10^{-4},
\label{bsgexptvalue:ref}
\eeq
where four independent measurements~\cite{exp} are combined, 
including a recent result from BaBar collaboration. 
\footnote{We add up all the errors of a single measurement 
in quadrature choosing a conservative value from asymmetric errors.}
Theoretical calculation of the inclusive decay
 in the SM has improved over several stages
 and next--to--leading order (NLO) QCD
 calculation has completed recently~\cite{bcmu02},
\beq
\brbsgamma = ( 3.70 \pm 0.30 ) \times 10^{-4}.
\eeq
There are a variety of numbers quoted by different authors
 possibly because of their choice of input, 
 however, they are overlapping with the measured value
 within their $1$ $\sigma$ errors, whose magnitude is similar to
 those of the experiments.
The SM again successfully explain the process with this level of accuracy
and therefore new physics effects should be confined
 to lie within the remaining uncertainty.
In particular, possible mass spectrum and flavor mixing
 of SUSY particles are expected to be strongly constrained
 from these results. To discuss them with the quoted accuracy, 
 it becomes important to introduce (SUSY--) QCD corrections beyond LO.

Going beyond the SM, NLO calculation has also been completed
 in two Higgs doublet model (2HDM)~\cite{2hdnlo,cdgg98}.
In supersymmetry, NLO formulas are available
only for limited parameter space~\cite{cdgg98susy}, while NLO-level analysis has been performed assuming minimal flavor violation (MFV)~\cite{large_tanb}.
In the analysis,
 two dominant effects, resummation of QCD logarithms and
 threshold correction to the bottom quark mass~\cite{bottomcorr}
 have been identified.
On the other hand, most of the analysis including general flavor mixing (GFM)
 in squark sector so far has been done using LO formulas
 or, at most, with LO matching of SUSY contribution~\cite{ggms96,
gluino,ekrww02}.

In this paper, we present the results of our analysis of $\bsgamma$
 in the minimal supersymmetric standard model (MSSM) with GFM in the squark sector.
In addition to the NLO contributions from
 the SM and the 2HDM,
we introduce NLO QCD corrections and leading NLO SUSY--QCD effects
 to dominant SUSY contributions, which enable us to compare the results
 with the experiments and existing NLO-level analysis in MFV. 

\section{Minimal Flavor Violation}

In the SM, after integrating out $W$ and top quark,
the short-distance physics is described by the following effective
 Lagrangian~\cite{gm01},
\beq
{\cal L} = \frac{4 G_F}{\sqrt{2}} V_{ts}^*V_{tb}
 \sum^8_{i=1} \left[ \Ci(\overline{\mu}){\it P}_{i} + \Cip(\overline{\mu}){\it P}_{i}^{\prime} \right],
\label{sm_operators:ref}
\eeq
where ${\it P}_i^{\prime}$ represent chirality partners of ${\it P}_i$,
 which are suppressed by $m_s/m_b$ in the SM while this is not
necessarily true in its extensions.
At $\overline{\mu}=\muw$, the Wilson coefficient
of the four-fermion operator $C_2$ has a tree level contribution
and dominant loop contribution comes to the coefficient of (chromo--) magnetic operator,
 $C_7$ ($C_8$).
These coefficients mix with each other by renormalization group evolution
 between $\muw$ and $\mub$. Then $\brbsgamma$ is evaluated at $\mub$ 
 based on the heavy--quark expansion.
For this part of calculation, we follow the NLO calculation of \cite{gm01}
with updated numbers in \cite{bcmu02}.

In the 2HDM, the charged Higgs boson $H^{-}$ also contributes
to $C_{7,8}$ replacing $W$ in the SM. 
This always adds constructively to the contribution from the SM,
 which pushes up the mass of $H^{-}$ until it almost decouples form
the process unless other contributions cancel it. 
We use matching condition in \cite{bmu00} at $\muw$
 to estimate this contribution at NLO. 

In the MSSM,  
 SUSY doubles the number of physical states
 and the contributions to $\bsgamma$ of CKM origin.
SUSY partners of $W$ and $H^{-}$ mix with each other and form
 charginos,
 which contribute to the process with up-type squarks as the corresponding
 contributions in the SM/2HDM. 
This contribution comes constructively or destructively depending on
 whether the higgsino mass parameter $\mu<0$ or $\mu>0$.
If we assume the CKM matrix is a unique source of flavor mixing in both
 quark and squark and limit ourselves to the above
 contributions (MFV framework), $\bsgamma$ provides a strict bound on
 mass spectrum of SUSY partners. 

\begin{figure}[t]
\vspace*{-0.4in}
\hspace*{-.70in}
\begin{center}
\begin{minipage}{9.2cm} 
\centerline{
\psfig{figure=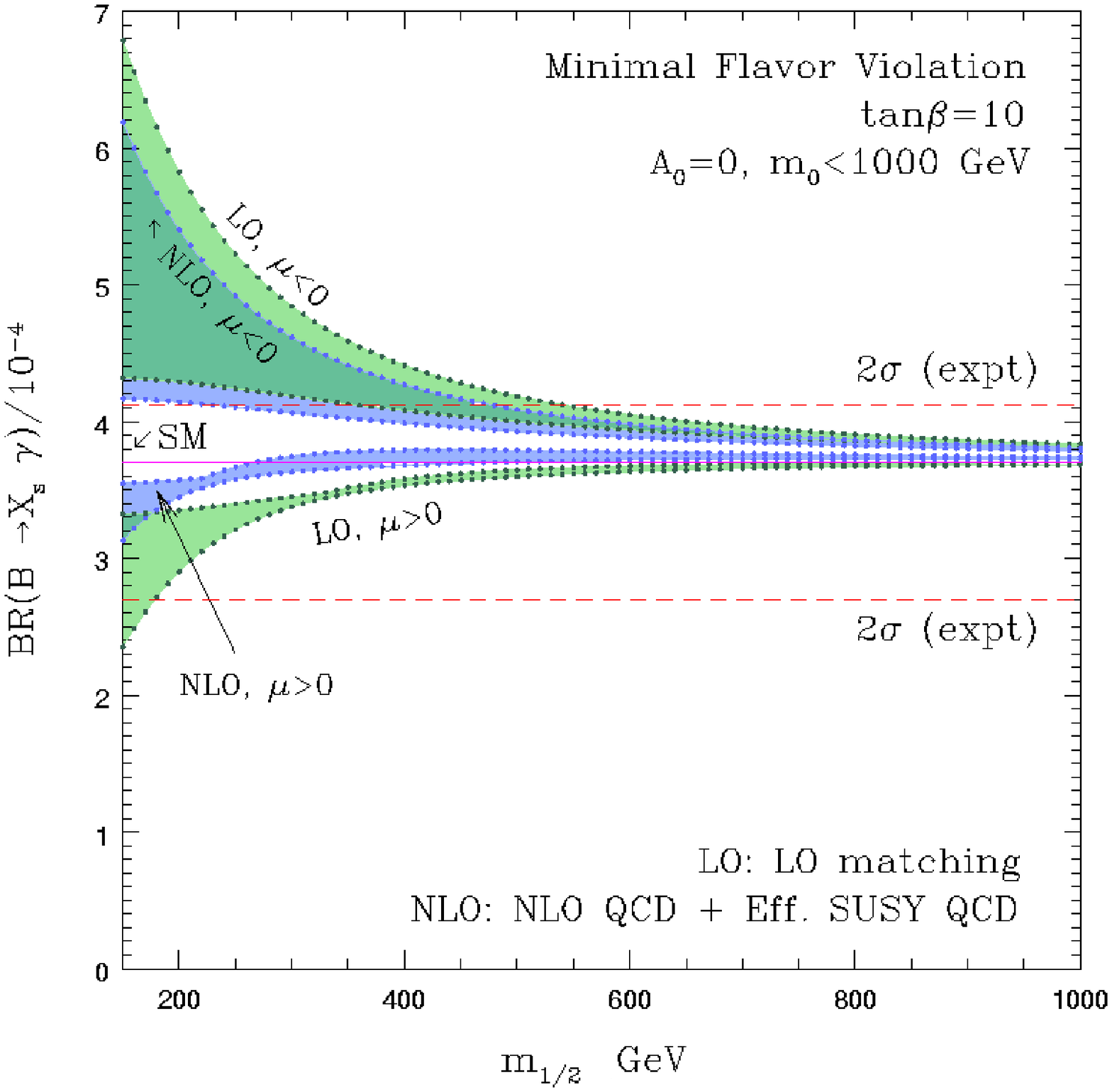,angle=0,width=4.6cm}
~~~~~~~~~~\psfig{figure=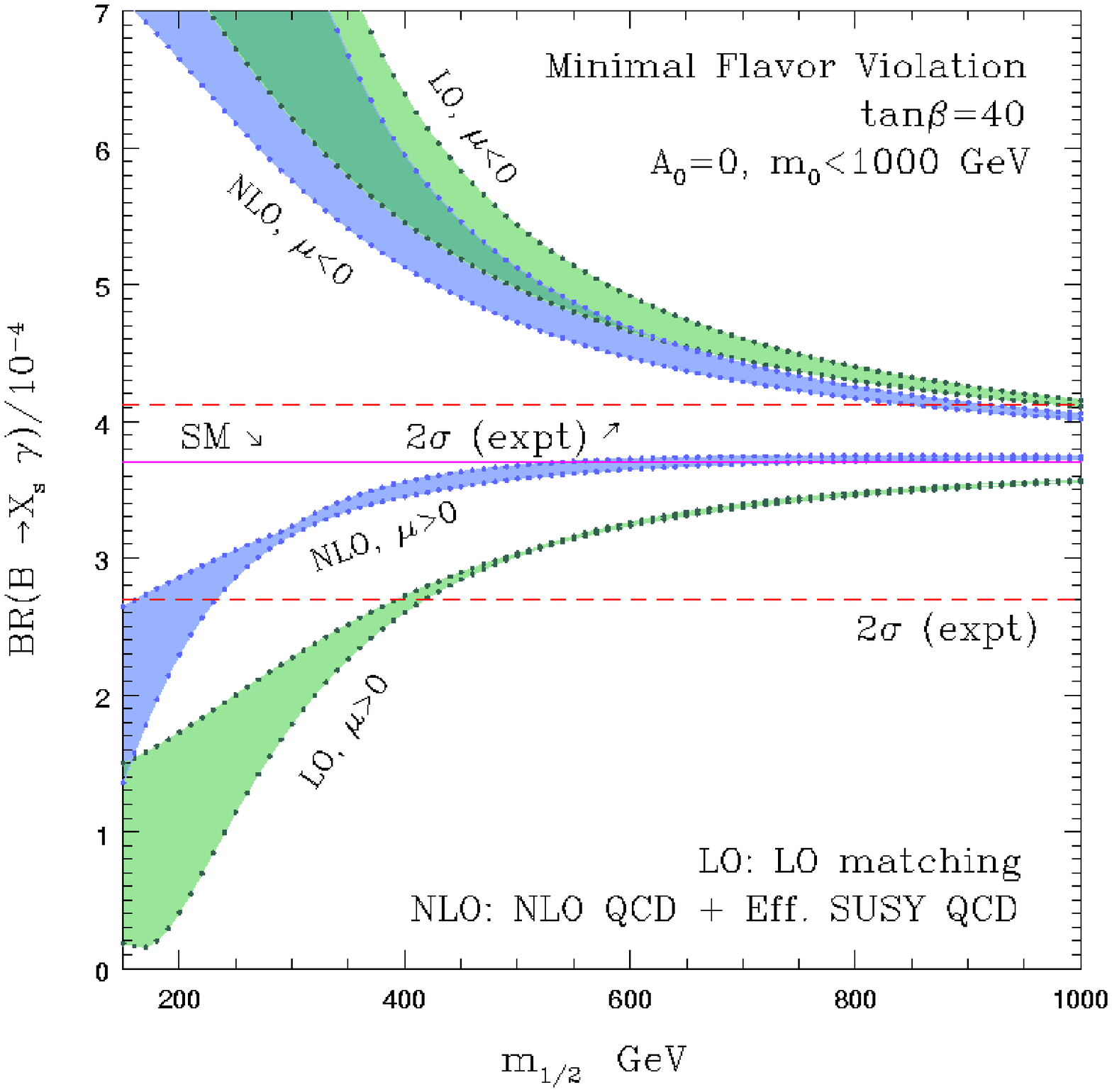,angle=0,width=4.6cm}}
\end{minipage}
\caption{\label{mfv:fig} {\small We plot $\brbsgamma$
\vs\ $\mhalf$ in the Constrained MSSM (MFV) for $\tan\beta = 10$ (left)
 and $\tan\beta = 40$ (right). 
We show the bands obtained by varying $\mzero$ in
the approximation of using only LO matching conditions (green) and of
including NLO corrections (purple). The SM prediction and $2\sigma$~CL
experimental limits are also marked.
}
}
\end{center}
\end{figure}
%
We illustrate this in Constrained MSSM (CMSSM). 
In FIG.~\ref{mfv:fig}, we take $\tan\beta
 = 10$ (left) and $\tan\beta=40$ (right) and fix $A_0=0$.
We scan $m_0$ over the region below $1000$ GeV and impose
the constraints from direct SUSY searches.
The green band shows a calculation, where the one--loop chargino
 contribution to $C_{7,8}$ is estimated at $\muw$. 
The blue band shows a improved calculation, where the SUSY contribution
 is estimated at the common mass scale of squarks, $\mususy$ using NLO QCD
matching condition in \cite{bmu00} and evolved down to $\muw$ using
NLO anomalous dimensions in \cite{misiakmunz95} with 6 quark flavor. 
Here we also take into account the effect of threshold correction 
to the bottom quark mass
 in the charged Higgs and higgsino couplings as described later, which 
 corresponds to
 a leading $\tan\beta$ enhanced part of the NLO SUSY--QCD correction.
This calculation reproduces well the results of 
 the second reference of \cite{large_tanb} in MFV.
The dashed two lines indicate the $2 \sigma$ experimental bounds and
the solid line
shows the prediction in the SM. 
As clearly seen, the $\mu$ negative case and
 low $\mhalf$ region are disfavored, especially, at larger value of $\tan
 \beta$.
Note that the NLO--level corrections are playing a crucial role in 
setting a lower bound of $\mhalf$ from the plot at $\tan\beta = 40$
 and $\mu>0$.

These bounds in MFV are rather model dependent, however,
 often used as one of reliable constraint to estimate other processes
in supersymmetry, especially, 
where flavor mixing is less important, such as
in detection of LSP cold dark matter~\cite{kim}.
These bounds often exclude parameter regions, which otherwise predict
 interesting signature in on--going or future experiments.
If we take this seriously, a question arises
 how robust the bounds really are, if we go beyond MFV.

\vspace{0.15cm}
\section{General Flavor Mixing}

Without any ad--hoc ansatz, gauge symmetry of
 the SM can not fix the flavor mixing
 in the soft SUSY breaking terms or, in turn, the squark mass matrices.
This general flavor mixing in down-type squarks induces 
new neutralino and gluino contributions to $\bsgamma$.
The latter is important because of enhancement
 by strong coupling constant.
To parameterize this new flavor
 mixing, it is convenient to introduce following dimensionless parameters,
\beq
\deltadll= \frac{(m_Q^2)_{23}}{\sqrt{(m_Q^2)_{22} (m_Q^2)_{33}}},~~~
\deltadlr= \frac{m_b(A_d)_{32}^*}{\sqrt{(m_D^2)_{33}(m_Q^2)_{22}}},
\label{gfm:eq}
\eeq
and similarly for $\deltadrr$ and $\deltadrl$.
Counter--parts of these flavor violating (FV) terms between
 $1$--$2$ generations are
 severely constrained by CP violating parameter in $K$--$\overline{K}$ 
mixing, while these $2$--$3$ FV--terms
 have been rather stringently
constrained by $\brbsgamma$~\cite{ggms96,gluino}. 

In this article
we extend the analyses including leading NLO corrections.
As explained in the previous section with MFV, we introduce NLO resummation
 of QCD logs to $C_{7,8}^{(\prime)}$ assuming the common mass scale, $\mususy$
 for squarks and gluinos, which tend to be heavier than other
states in most of interesting SUSY breaking models.
 As with the chargino contribution,
 we follow \cite{bmu00} for NLO QCD matching
 of the neutralino and gluino contributions. 
We checked that SUSY loop--induced contributions to $C_{1-6}^{(\prime)}$ is
 numerically less important in the estimation of $\brbsgamma$.
\footnote{Detailed analysis at LO~\cite{gluino} shows 
that a complete analysis with GFM introduces
 new operators other than those in Eq.(\ref{sm_operators:ref}).
Inclusion of them is out of scope of this paper, while the analysis 
has shown that they are less important at LO.}
NLO matching also requires ${\it O}(\alpha_s)$ corrections
 from gluino. 
To include this NLO SUSY--QCD corrections, we introduce
 the effective coupling constants for $W$ and $H^-$
 after decoupling of squarks and gluino~\cite{cdgg98}.
We have re-calculated them including full flavor mixing. 
On the other hand, this procedure is not justified
 for the SUSY contributions themselves which 'decouple' in its framework.
Instead of calculating full two-loop diagrams,
we take into account leading $\tan\beta$ enhanced parts of them.
We insert finite counter--terms into the LO SUSY contributions,
 which are required by the renormalizaton of the quark mass matrix
 to keep their flavor--diagonal form.
These 'two--loop' level corrections are enhanced by an explicit
$\tan\beta$ factor relative to the corresponding two--loop diagrams. 
This insertion is performed on higgsino coupling~\cite{large_tanb}
 and the F--term contributions of squark mass matrix.
In the calculation, we include all sources of flavor mixing by 
 numerically diagonalizing mass matrices,
 instead of using the mass--insertion approximation. 
Technical details of the calculation can be
 found in the reference~\cite{or1}.
In the following section, we will illustrate the impact of these
 corrections beyond LO and robustness of the MFV prediction against
 GFM with them.

\section{Focusing effect}

We have identified NLO focusing effect especially in large
 $\tan\beta$ and $\mu>0$ region, which considerably relaxes 
the constraints on the squark flavor--mixing.
We illustrate this in CMSSM inserting the FV--terms at $\mususy$.
%
\begin{figure}[h]
\vspace*{-0.3in}	
\begin{center}
\begin{minipage}{13.8cm} 
\centerline
{\hspace*{-.2cm}\psfig{figure=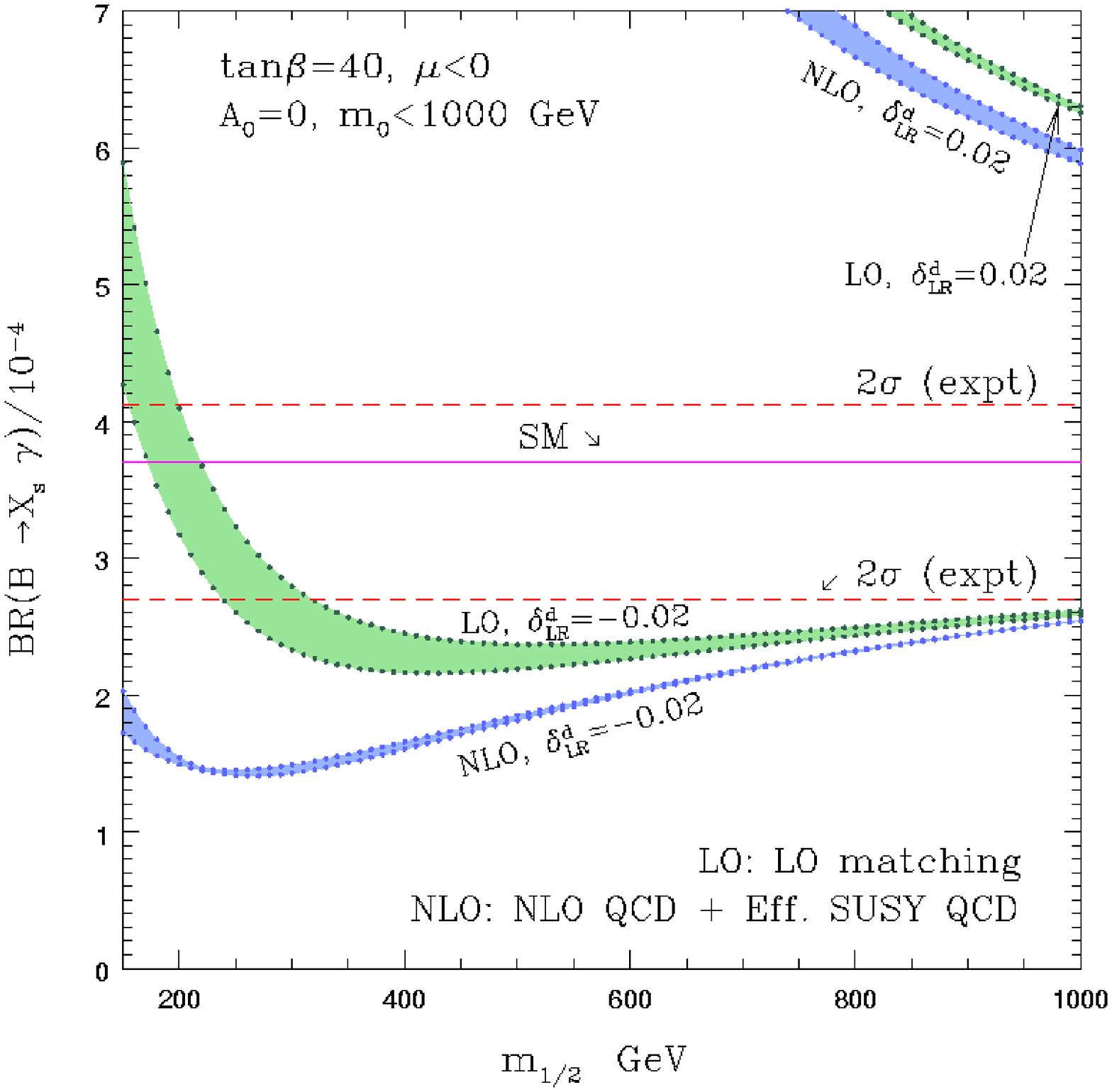, angle=0,width=4.6cm} 
	    \hspace*{-.3cm}\psfig{figure=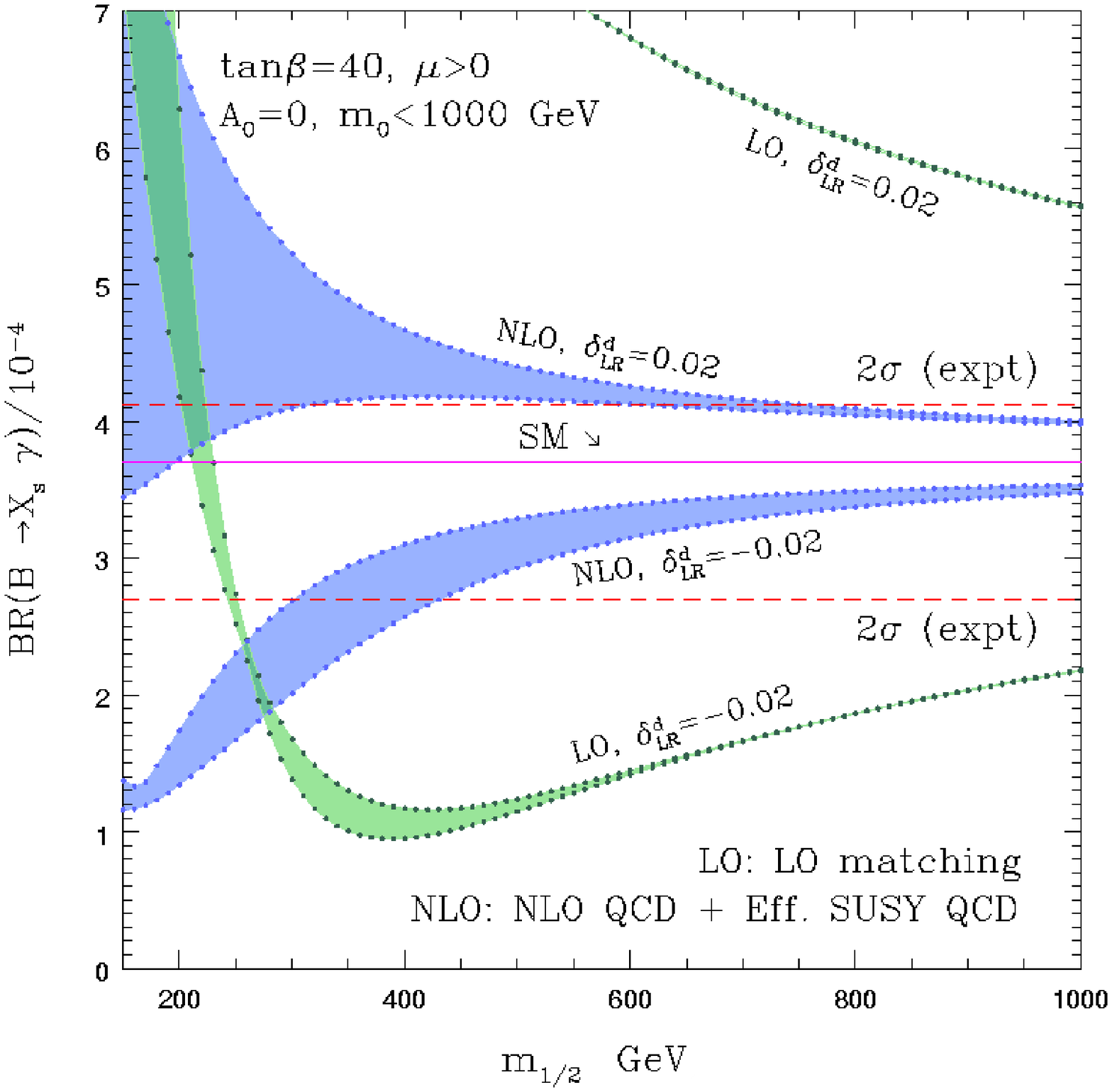, angle=0,width=4.6cm}
	    \hspace*{-.3cm}\psfig{figure=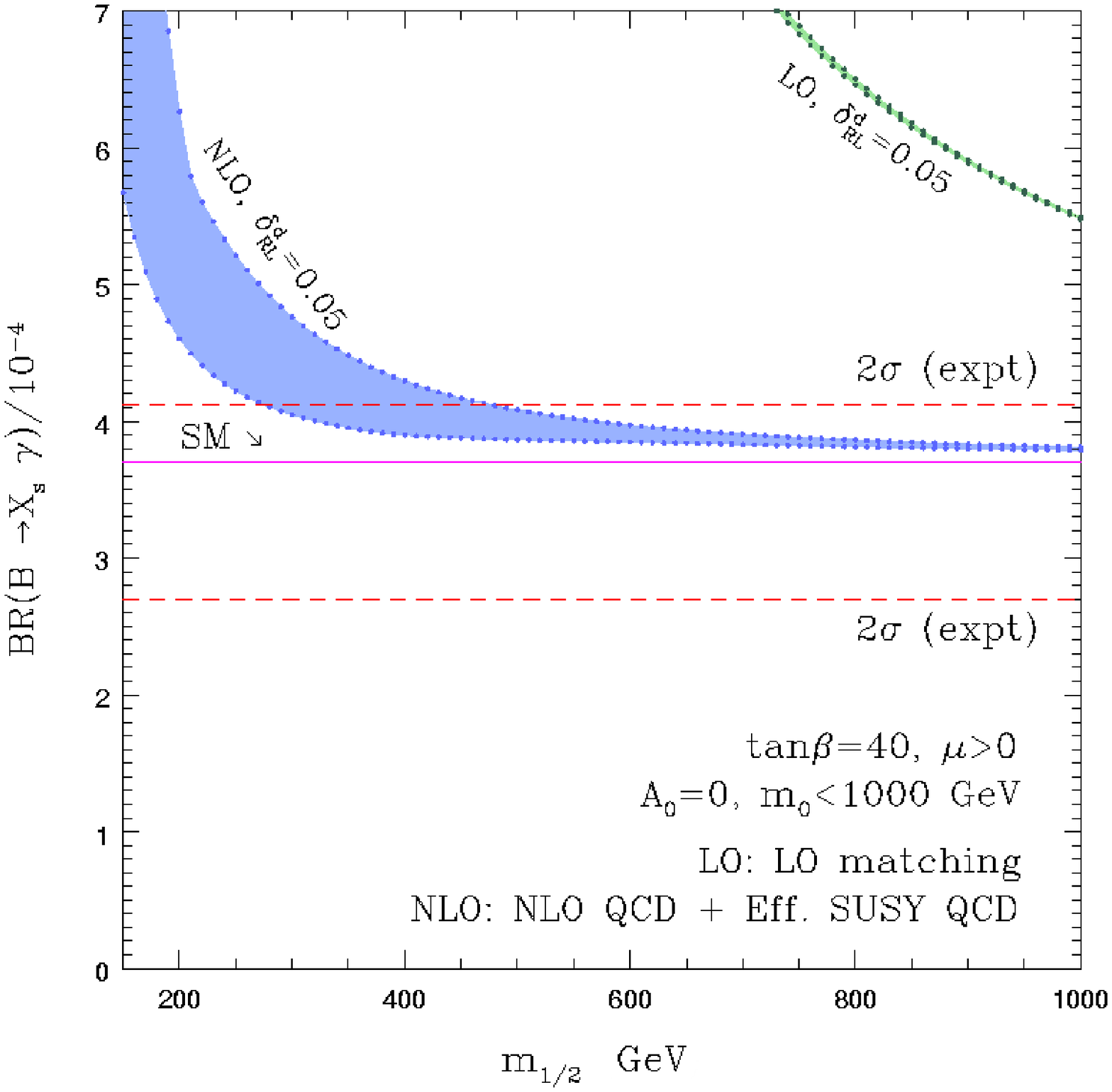, angle=0,width=4.6cm}
}
\end{minipage}
\end{center}
\begin{center}
\begin{minipage}{13.8cm} 
\centerline
{\hspace*{-.2cm}\psfig{figure=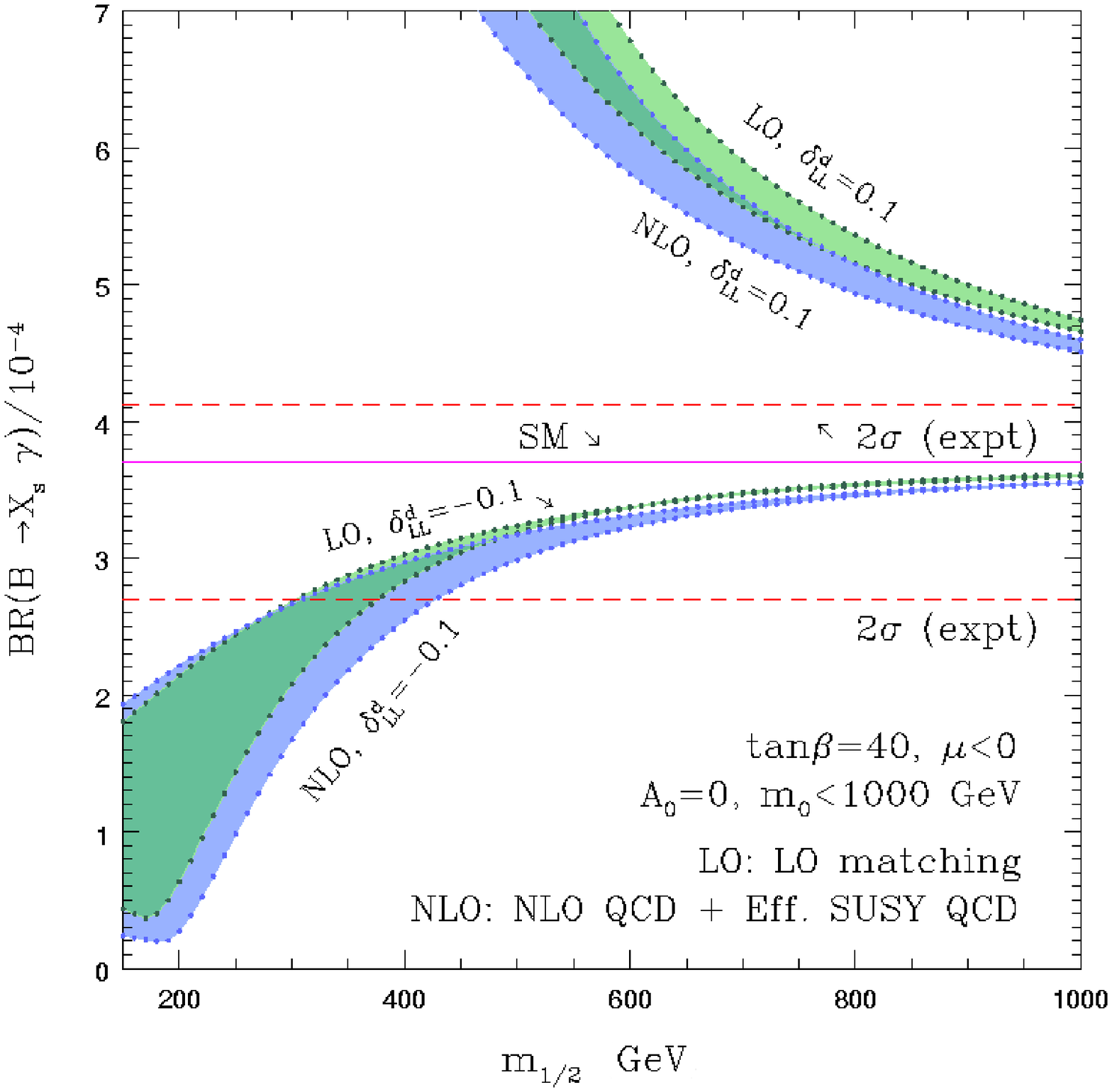, angle=0,width=4.6cm} 
	    \hspace*{-.3cm}\psfig{figure=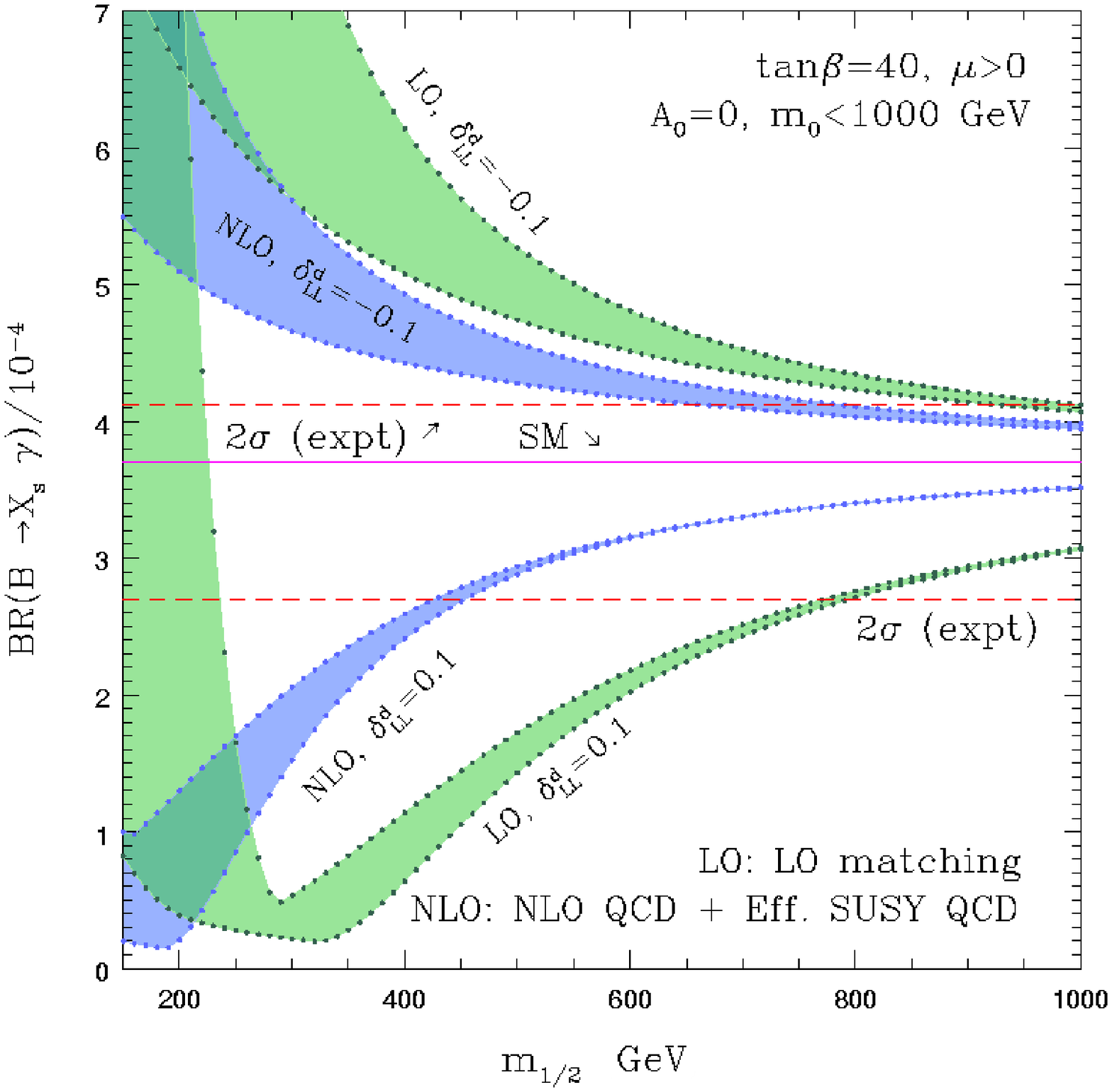, angle=0,width=4.6cm}
	    \hspace*{-.3cm}\psfig{figure=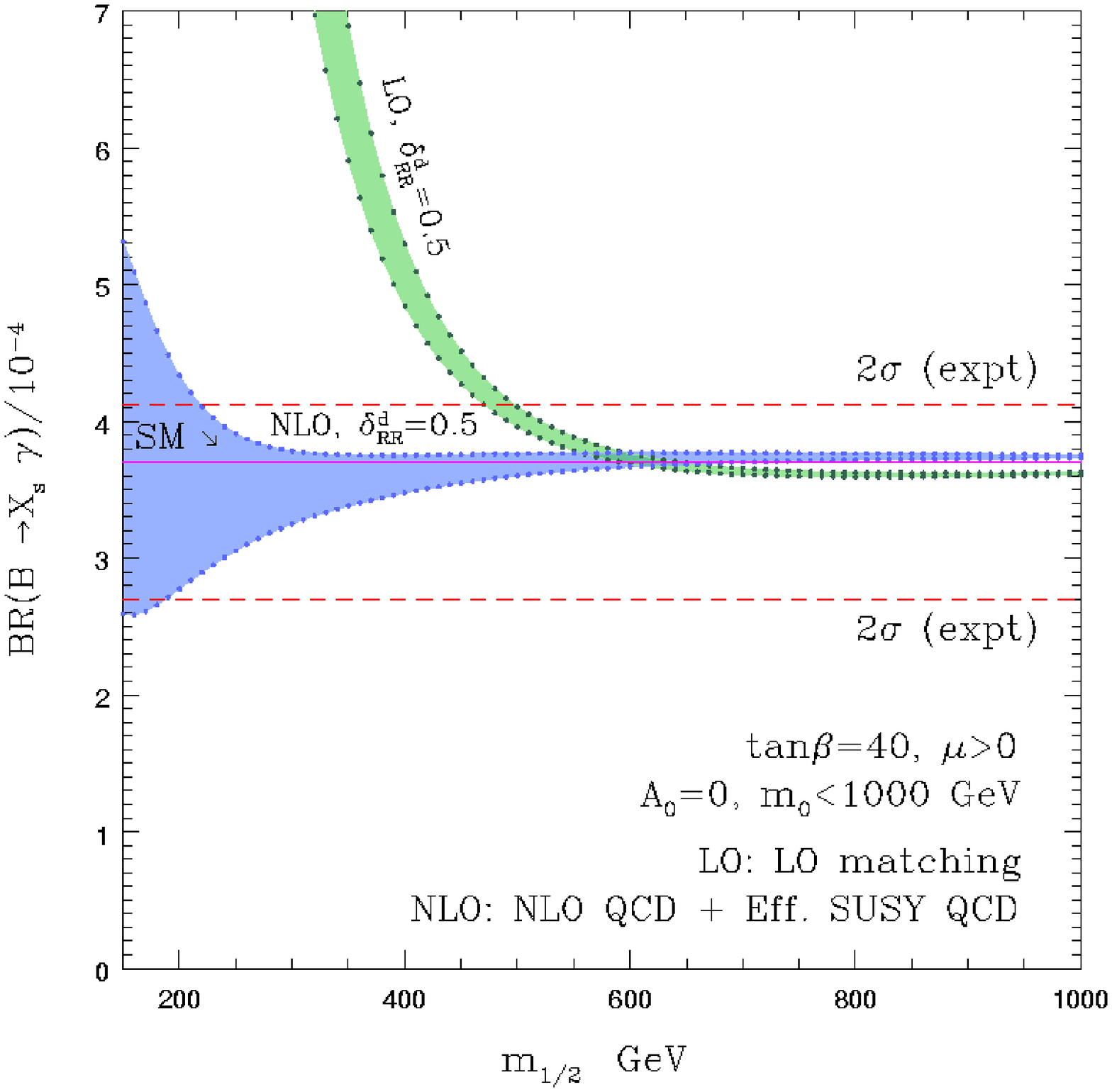, angle=0,width=4.6cm}
}
\end{minipage}
\caption{\label{brbsg-delta:fig} {\small
Similar plots as in FIG.~\ref{mfv:fig}
 with General Flavor Mixing in down-type squark.
}}
\end{center}
\end{figure}
%
FIG.~\ref{brbsg-delta:fig} shows similar plots  
 of $\brbsgamma$ as FIG.~\ref{mfv:fig} with GFM for $\tan\beta=40$. 
The upper middle window is a case for $\deltadlr=\pm 0.02$ at $\mu>0$.
The green bands correspond to a calculation with
 LO matching condition and the blue bands show
 that with the leading NLO corrections. 
The branching ratio is strongly focusing toward its SM value,
once the NLO-level corrections are included.  
Similar strong focusing at $\mu>0$ is generic for $\deltadrl=0.05$ (upper
right), $\deltadll=\pm 0.1$ (lower middle) and $\deltadrr=0.5$ (lower right).
We do not plot the case for $\deltadrl=-0.05$ and $\deltadrr=-0.5$
 because the branching ratio is almost symmetric about
 these FV--terms.
At $\mu<0$, focusing is not obvious as seen in the cases of $\deltadlr$ (upper left) and $\deltadll$ (lower left) 
and we suppress $\deltadrl$ and $\deltadrr$ cases here.

Focusing effect comes from two sources.
Firstly,  renormalization group evolution of $C_{7,8}$ from $\mususy$ to
$\muw$
 reduces the overall amplitude of these coefficients.
NLO QCD matching of them at $\mususy$ brings further suppression.
Secondly, gluino contribution to the mass matrix of down-type quark
 shows remarkable alignment with the dominant Wilson coefficients
 at $\mu>0$, which is taken into account by
 including the leading NLO SUSY--QCD corrections at $\mususy$
as described in the previous section. 
This considerably reduces the LO gluino contribution to $C_{7,8}^{(\prime)}$. 
This alignment is enhanced by $\tan\beta$ and dominates the focusing
 effect in the figure at $\mu>0$. 
At $\mu<0$, the SUSY--QCD corrections cause
anti--alignment instead, 
which competes with the overall suppression by renormalization group evolution
 and results in small focusing ( or de--focusing ) in the figure.
Some NLO suppression of SUSY contribution
 already exits in MFV as shown in FIG.~\ref{mfv:fig}, 
however, flavor mixing is essential for the focusing effect with GFM
 as described above.
More detailed discussion can be found in reference~\cite{or1}.


\vspace{0.15cm}
\noindent
\section{Relaxation of Constraints on SUSY and FV Terms}  

\begin{figure}[b]
\vspace*{-0.3in}	
\begin{center}
\begin{minipage}{13.8cm} 
\centerline
{\hspace*{-.2cm}\psfig{figure=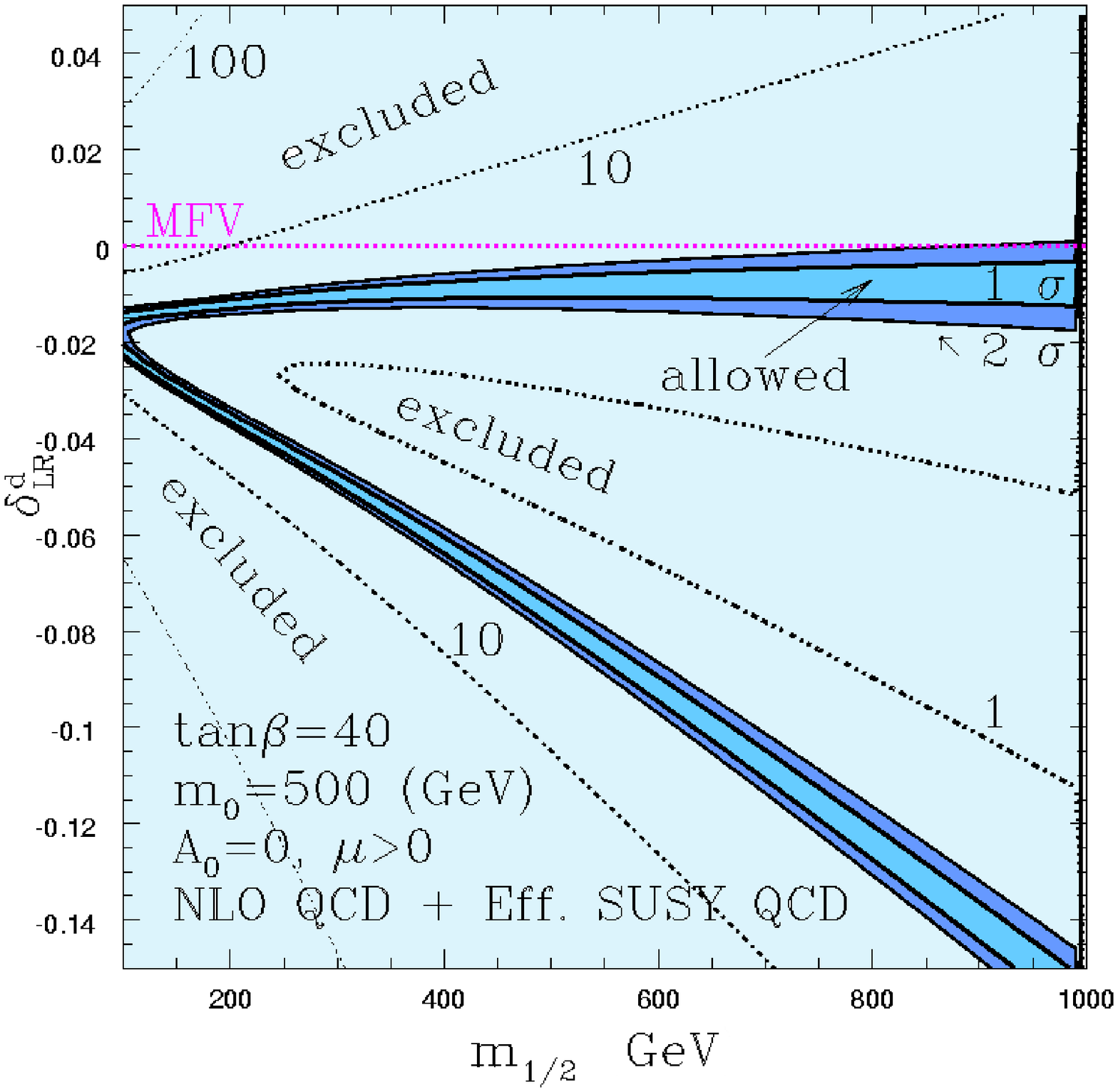, angle=0,width=4.6cm} 
	    \hspace*{-.3cm}\psfig{figure=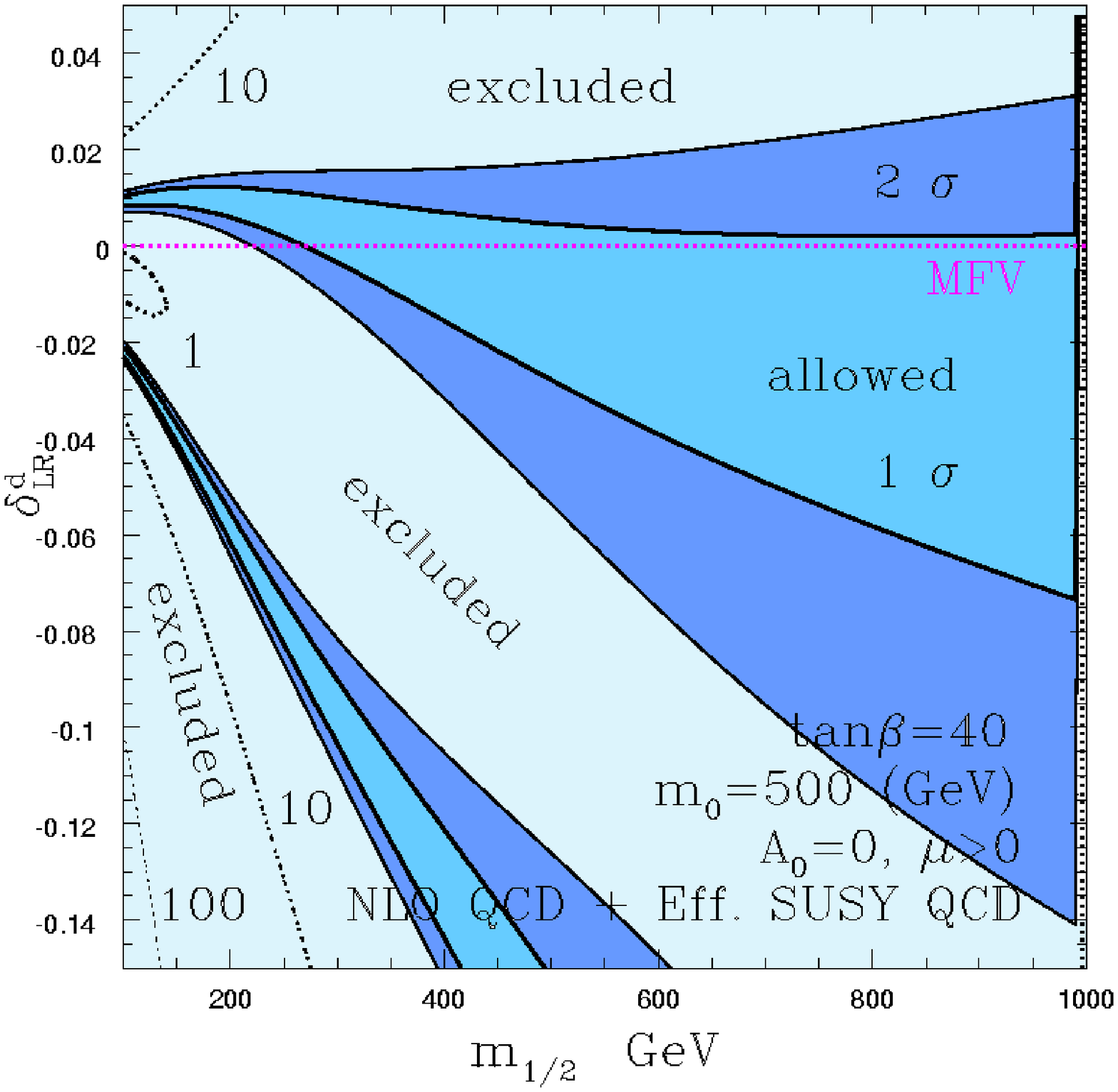, angle=0,width=4.6cm}
	    \hspace*{-.3cm}\psfig{figure=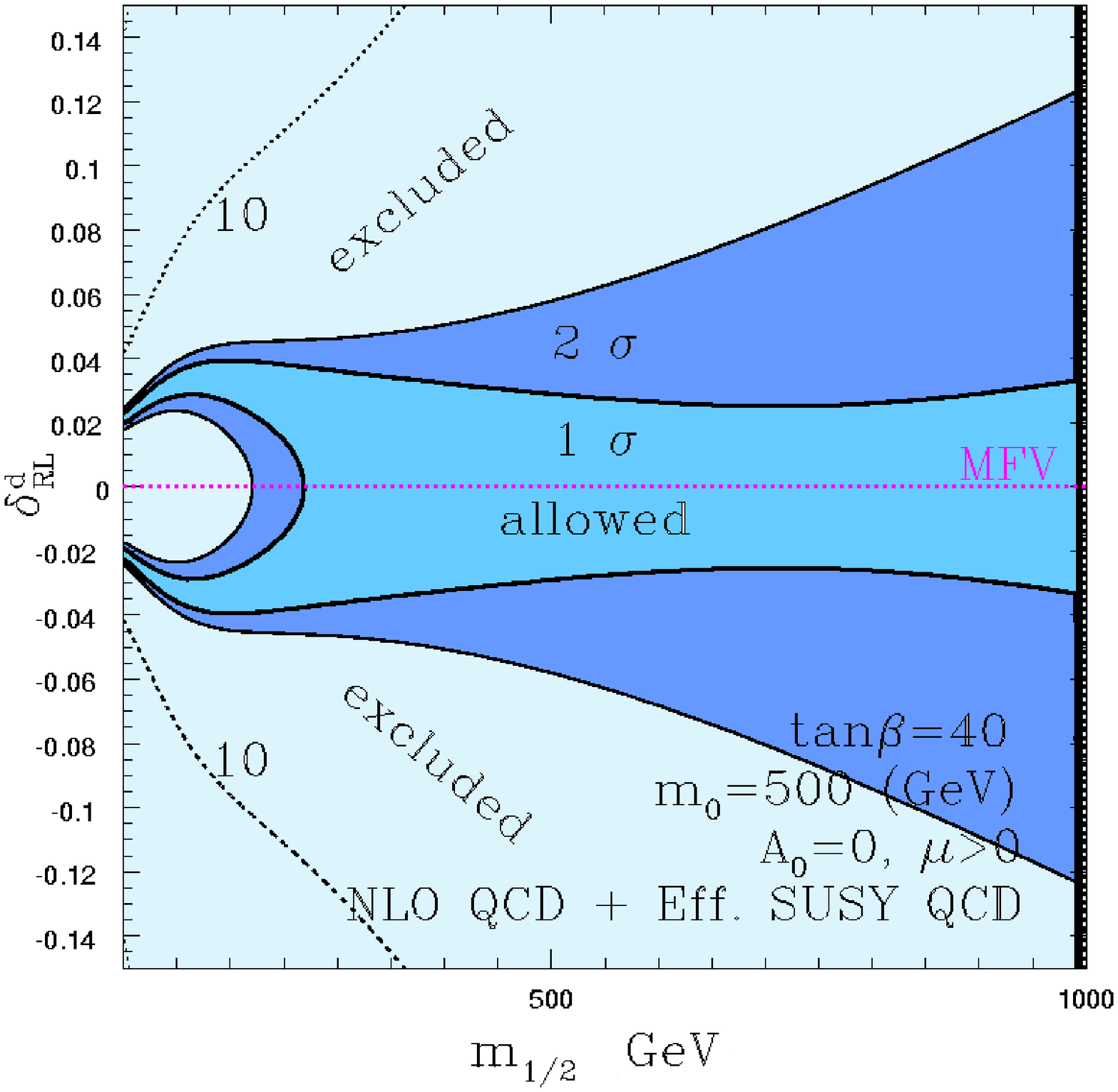, angle=0,width=4.6cm}
}
\end{minipage}
\end{center}
\begin{center}
\begin{minipage}{13.8cm} 
\centerline
{\hspace*{-.2cm}\psfig{figure=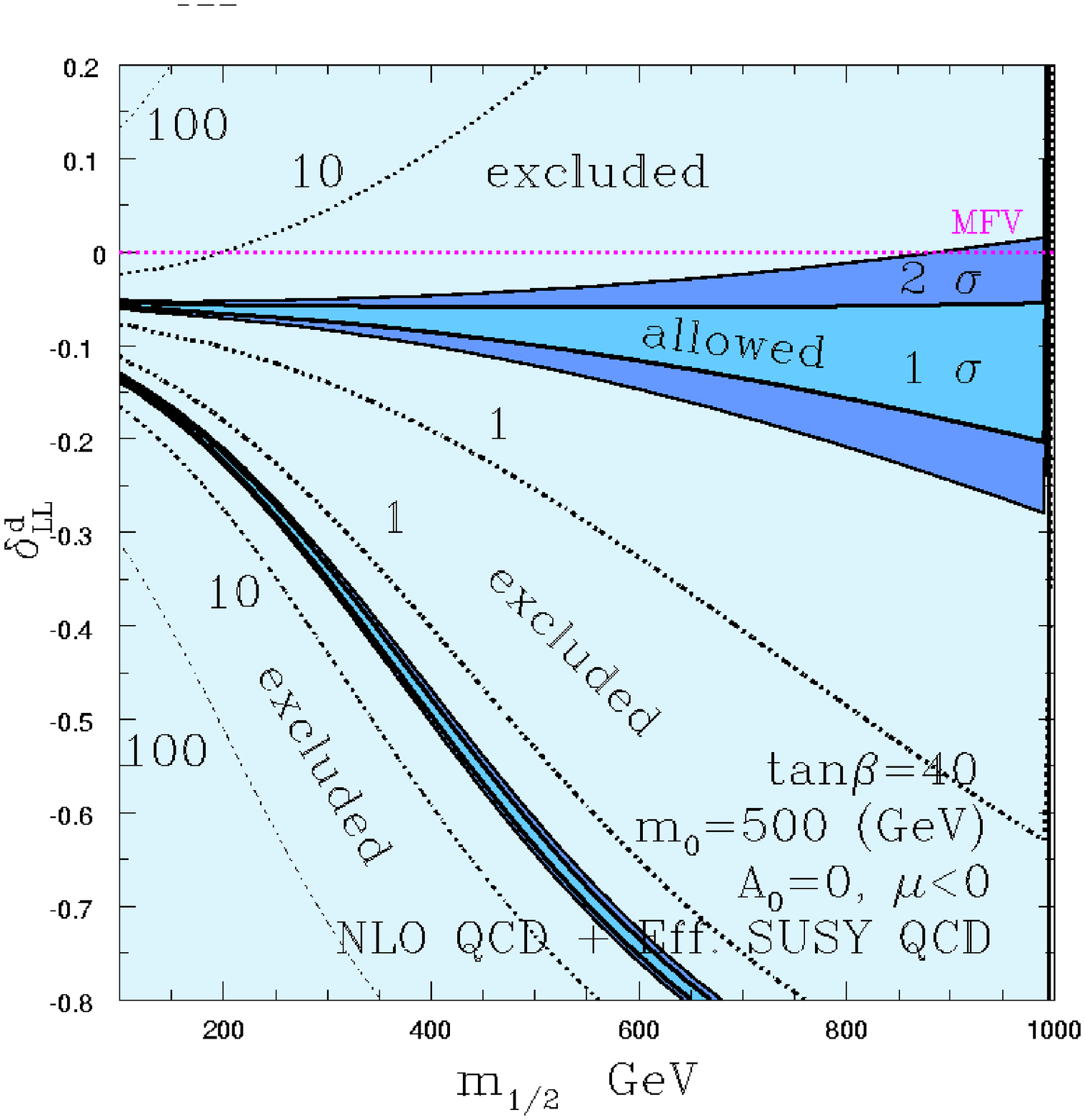, angle=0,width=4.6cm} 
	    \hspace*{-.3cm}\psfig{figure=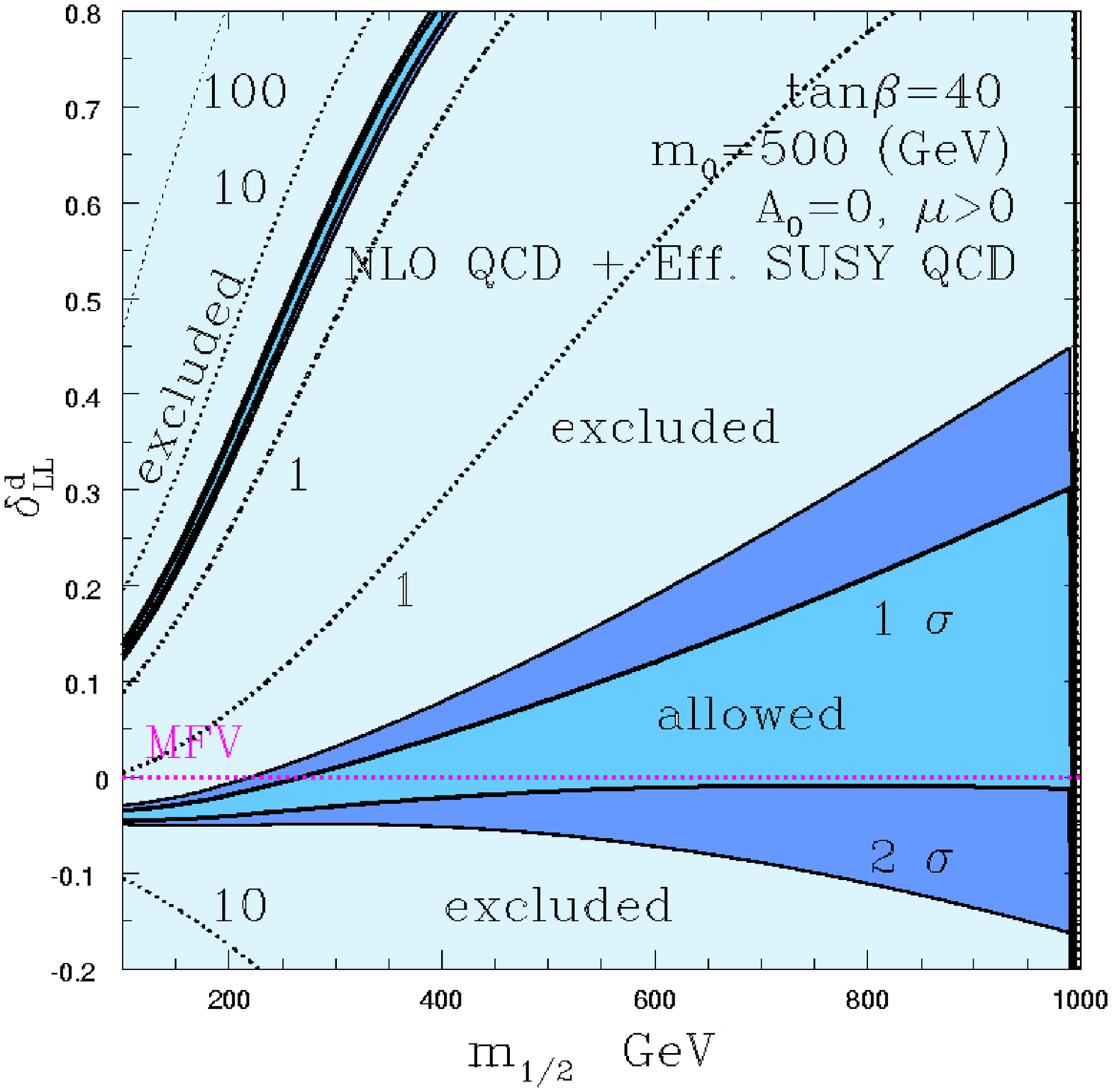, angle=0,width=4.6cm}
	    \hspace*{-.3cm}\psfig{figure=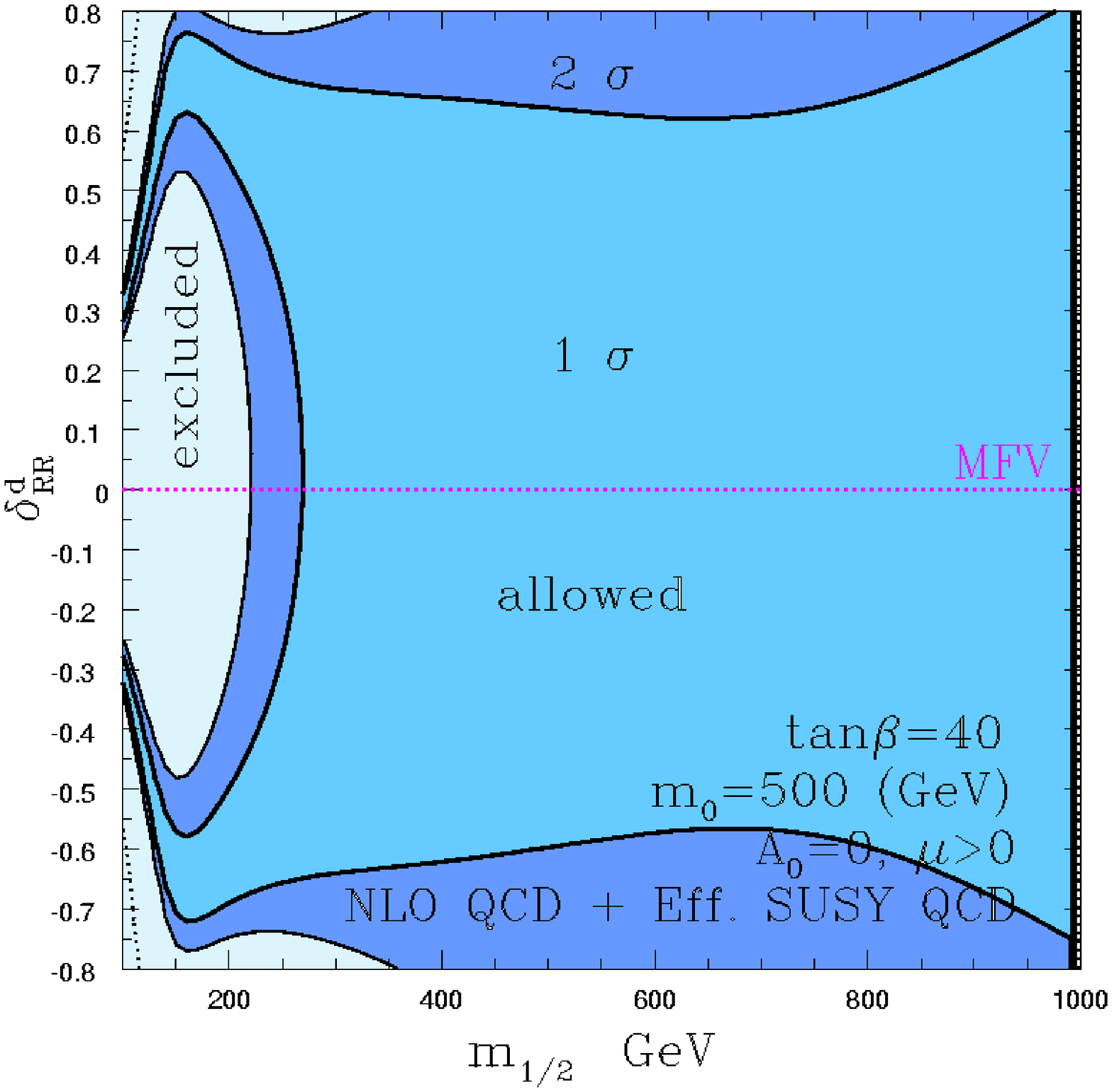, angle=0,width=4.6cm}
}
\end{minipage}
\caption{\label{contour:fig} {\small We plot contours of $\brbsgamma$
\vs\ $\mhalf$ and various flavor violating terms in the Constrained MSSM. 
Regions
beyond the $1\sigma$ (light blue) and $2\sigma$~CL (dark blue) agreement with
experiment (Eq.~(\ref{bsgexptvalue:ref})) are marked ``excluded''.
}}
\end{center}
\end{figure}

The introduction of GFM changes the whole picture of $\bsgamma$
 constraint on SUSY parameters.
Therefore, the focusing effect has strong implication
 on analysis of SUSY models beyond MFV.
We again illustrate this in CMSSM.
FIG.~\ref{contour:fig} shows contour--plots of $\brbsgamma$ in the
 $\mhalf$ and FV--term plane.
Here, we choose the CMSSM parameters as $\tan\beta=40$, $A_0=0$ and
 $m_0=500$ GeV.
The upper left (middle) window is for $\deltadlr$ at $\mu<0$ ($\mu>0$).
The light (dark) blue bands show allowed regions within the $1
\sigma$ ($2 \sigma$) experimental error. Outside of these regions is
 labeled as 'excluded'.
Comparing the plot at $\mu<0$, remarkably wider allowed--regions
 at $\mu>0$ roughly measure an impact of the focusing effect. 
It is worth to comment that the 'excluded' region between the two bands
 is very shallow and small reduction of experimental value
( for example, if we do not include the recent BaBar data )
 easily accommodates most of this region with $2 \sigma$ error
 except for small $\mhalf$ part. 
Similar NLO-level relaxations exist
 in the plots for $\deltadrl$ (upper right) and $\deltadrr$ (lower right).
These plots are almost symmetric about the FV--terms because new gluino
 contributions appear in $C_{7,8}^{\prime}$ and contribute to the branching
 ratio quadratically without interfering the MFV contributions in $C_{7,8}$.
In these cases, new contributions always come constructively and
 almost all of the region in the plot is not allowed at $\mu<0$
, which is omitted in the figure.
 In the case of $\deltadll$ ( lower left and middle windows ), 
the strong suppression of gluino contribution is rather unclear 
 because of a new chargino contribution generated
 by coexisting SU(2)--related GFM among up-type squarks.
In these plots, $\brbsgamma$ seems to be more sensitive
 to $\deltadlr$ and $\deltadrl$ than $\deltadll$ and $\deltadrr$. 
However, it should be noted that natural scale of $\deltadlr$ ($\deltadrl$) is suppressed
by $m_b/\msusy \lsim {\it O}(10^{-2})$ in Eq.~(\ref{gfm:eq}), where we follow the definitions
 widely used in the literature.
Even though the focusing effect considerably reduces the gluino
 contribution generated by GFM,
 the branching ratio still shows strong dependence on $\deltadll$  and
 $\deltadlr$ because of the interference between the gluino contributions
 and the MFV contributions to $C_{7,8}$.
In addition, the new chargino contribution is present
 in the $\deltadll$ case. 
We can read the lower bound of $\mhalf$ in MFV as $\mhalf \gsim 200$
 ($600$) GeV at $\mu>0$ ($\mu<0$). 
However, FV--term of $|\deltadll|\sim 0.05$ ( $|\deltadlr|\sim 0.01$ ) 
could easily remove (or double) this bound, which is not necessary considered
 as a fine tuning. 
This is more striking at $\mu<0$, where the focusing effect is not so obvious.
On the other hand, this bound does not change much with $\deltadrr$
 or $\deltadrl$ if $\deltadrr \lsim
 0.4$ and $\deltadrl \lsim 0.015$, where the focusing effect plays
 crucial role in relaxing the constraints. 
This means that NLO-level calculation is essential
 to test flavor models, 
for example, based on SO(10) SUSY GUT, which tend to have large $\large \tan\beta$.


\vspace{0.15cm}
\noindent
\section{Conclusions}
We have analysed $\brbsgamma$ in MSSM with general flavor mixing
including leading NLO corrections within the SM operator 
basis, assuming the common mass scale, $\mususy$ for squarks and gluino.
We have identified the NLO focusing effect, which reduces
 the gluino contribution to 
$\brbsgamma$ relative to that with LO matching condition.
This effect leads to a considerable relaxation of the constraints on flavor
 mixing terms, especially for $\deltadlr$.
Including the focusing effect, we have shown in CMSSM that
 $\brbsgamma$ is still very sensitive to $\deltadll$ and $\deltadlr$.
The lower bound for $\mhalf$ obtained in MFV 
can be easily removed by them even at $\mu<0$, 
while it is rather stable against $\deltadrr$ and $\deltadrl$.
NLO-level calculation appears to be essential in
 analysis of SUSY models beyond MFV with large $\tan\beta$. 
This calls for a future complete NLO calculation
 of $\brbsgamma$ in supersymmetry which
 could be applicable in more general circumstance.

\section*{Acknowledgments}
We would like to thank P.~Gambino and T.~Blazek for helpful comments.


\end{document}